# Parikh Images of Regular Languages: Complexity and Applications

Anthony Widjaja To

LFCS, School of Informatics, University of Edinburgh

**Abstract**

*We show that the Parikh image of the language of an NFA with $n$ states over an alphabet of size $k$ can be described as a finite union of linear sets with at most $k$ generators and total size $2^{O(k^2 \log n)}$, i.e., polynomial for all fixed $k \geq 1$. Previously, it was not known whether the number of generators could be made independent of $n$, and best upper bounds on the total size were exponential in $n$. Furthermore, we give an algorithm for performing such a translation in time $2^{O(k^2 \log(kn))}$. Our proof exploits a previously unknown connection to the theory of convex sets, and establishes a normal form theorem for semilinear sets, which is of independent interests. To complement these results, we show that our upper bounds are tight and that the results cannot be extended to context-free languages. We give four applications: (1) a new polynomial fragment of integer programming, (2) precise complexity of membership for Parikh images of NFAs, (3) an answer to an open question about polynomial PAC-learnability of semilinear sets, and (4) an optimal algorithm for LTL model checking over discrete-timed reversal-bounded counter systems.*

## 1 Introduction

A *semilinear set* is any subset of $\mathbb{N}^k$ that can be described as a finite union of *linear sets* over $\mathbb{N}^k$ of the form $\{\mathbf{v}_0 + \Sigma_{i=1}^m a_i \mathbf{v}_i : a_1, \ldots, a_m \in \mathbb{N}\}$ for some *offset* $\mathbf{v}_0 \in \mathbb{N}^k$ and *generators* $\mathbf{v}_1, \ldots, \mathbf{v}_m \in \mathbb{N}^k$. The well-known Parikh's Theorem states that semilinear sets are effectively equivalent with the sets of letter-counts (a.k.a. Parikh images) of regular languages and those of context-free languages [23]. This theorem is well-known to be a fundamental result in automata theory with a plethora of applications. These include verification [7, 13, 16, 30], automata and logics over unranked trees with counting [26], equational horn clauses [32], and word-automata theory itself [5, 14, 20], among many others. Most practically-motivated applications (e.g. [7, 13, 14, 16, 20, 26, 30, 32]), however, require more than the effective equivalence of such representations. The issues of succinctness and complexity of translations among different representations are also equally important. These issues are the main subject of this paper, where nondeterministic finite automata (NFA) and context-free grammars (CFG) are adopted as representations of regular languages and context-free languages (respectively). [As we shall see, our results also hold for other standard representations of regular and context-free languages.]

There is a trivial polynomial-time translation from a given semilinear set $S$ (where numbers are given in unary) to an NFA or a CFG whose Parikh image represents $S$. On the other hand, the reverse (more important) direction is not yet fully understood. All known translations from NFA and CFG to semilinear sets (e.g. see [10, 18, 23, 26] and the references therein) yield at least *exponentially many* linear sets. It was not clear whether (and perhaps, to what extent) such an exponential blow-up can be avoided.

Some partial answers are available. Chrobak-Martinez Theorem[1] [6, 20] shows that, given an NFA with $n$ states over a unary alphabet (i.e. with one letter), one could compute in poly-time a union of $O(n^2)$ many arithmetic progressions $\{a + tb : t \in \mathbb{N}\}$ such that $a = O(n^2)$ and $b = O(n)$. This theorem has been applied to derive optimal algorithms for other problems in automata theory (e.g. see [14, 20]) and, recently, in the verification of *one-counter systems* (e.g. see [13]). Note that arithmetic progressions are simply linear sets with exactly one generator. Chrobak-Martinez Theorem is in stark constrast with the known general translations from NFA (e.g. [10, 18, 23, 26, 32] and the references therein), which produce a union of exponentially many linear sets with unbounded number of generators *even over unary alphabet*. On the other hand, it was shown in [24] that a CFG $G$ over unary alphabet $\{a\}$ could be used to encode the language $\{a^{2^n}\}$, where $n$ is roughly the size of $G$. Therefore, at least without allowing binary representation in the output, the size of the semilinear sets could be exponential in the size of $G$.

Chrobak-Martinez Theorem suggests one obvious generalization: given an NFA with $n$ states over an alphabet of any fixed size $k \geq 1$, one could construct in poly-time a union of polynomially many linear sets with at most $k$ gen-erators whose offsets and generators contain only numbers of size (in unary) polynomial in $n$. Such a generalization would be interesting for two reasons. First, the size of the alphabet is often much smaller than the number of states in the NFA. Second, Lenstra [19] has given a poly-time algorithm for integer programming over any fixed number of variables. As we shall see, this would yield better complexity in applications requiring the use of Parikh's Theorem.

---

[1]Unfortunately, their proofs contain a subtle error, which were only recently fixed in [29]

One immediate hurdle in proving this general version of Chrobak-Martinez Theorem is that it was not known whether for every fixed $k \geq 1$ there exists a poly-time algorithm, which given a semilinear set in $\mathbb{N}^k$ (in unary), outputs an equivalent finite union of linear sets with at most $k$ generators and total size polynomial in the given semilinear set. For $k = 1$, this is simply a corollary of solutions to the *Frobenius problem* [25]. For $k = 2$, this has been proved by Abe [1]. However, his proof makes use of geometric facts in $\mathbb{R}^2$ that are not available in higher dimensions. In general, it was not known whether the weaker statement requiring only *existence* actually hold for all fixed $k \geq 1$.

**Contributions** We prove the general version of Chrobak-Martinez Theorem: given an NFA with $n$ states over an alphabet of size $k$, we can compute in time $2^{O(k^2 \log(kn))}$ a finite union of linear sets with at most $k$ generators and total size $2^{O(k^2 \log n)}$ (even when unary representations of numbers are imposed). To this end, we establish a *normal form theorem* for semilinear sets: any semilinear set in $\mathbb{N}^k$ of size $n$ (under unary representation of numbers) could be converted into an equivalent union of $2^{O(k \log(kn))}$ linear sets $\{\mathbf{v}_0 + \Sigma_{i=1}^k a_i \mathbf{v}_i : a_i \in \mathbb{N}\}$, where numbers in $\mathbf{v}_0$ (resp. $\{\mathbf{v}_i\}_{i=1}^k$) cannot exceed $2^{O(k \log(kn))}$ (resp. $O(n)$). Furthermore, we show that this conversion can be performed in time $2^{O(k^2 \log(kn))}$. In fact, a similar result is shown to hold for semilinear sets over $\mathbb{Z}$ (i.e. where offsets and generators are in $\mathbb{Z}^k$). The proof of the normal form theorem makes use of the well-known *Caratheodory's theorem* [33] from convex geometry. Our normal form theorem for semilinear sets are of independent interests and have applications beyond the computation of Parikh images of NFAs.

To complement our upper bounds, we show that for every fixed $k \geq 1$ there exist infinitely many NFAs $\{A_n\}$ over an alphabet of size $k$ such that $A_n$ has $n$ states and its Parikh image contains $\Omega(n^{k-1})$ linear sets. This implies that we cannot remove $k$ from the exponent of the running time of translations from NFA to semilinear sets. Furthermore, we give infinitely many CFG over a unary alphabet whose Parikh images contain at least $2^n$ linear sets, where $n$ is roughly the size of the CFG, and thus strengthening the initial lower bound from [24].

We give a few applications of our main results: (1) a new polynomial fragment of integer programming, (2) precise complexity of membership for Parikh images of NFAs, (3) an answer to an open question posed by Abe [1] about polynomial PAC-learnability of semilinear sets, and (4) an optimal algorithm for LTL model checking over discrete-timed reversal-bounded counter systems,

Finally, since there is a poly-time translation from regular expressions to NFAs (e.g. see [27]), our upper bound result for NFAs transfer directly to regular expressions. Although NFAs are more succinct than regular expressions [11] and deterministic automata, our lower bound result also transfers to regular expressions and deterministic automata. Furthermore, since CFGs and pushdown automata are poly-time equivalent (e.g. see [27]), our result for CFGs directly transfer to pushdown automata.

**Related work** It is of course well-known that semilinear sets also coincide with subsets of $\mathbb{N}^k$ expressible in Presburger arithmetic [12] (i.e., first-order logic over $(\mathbb{N}, +)$). Such an alternative representation has indeed been fruitfully exploited. In fact, using a technique developed by Esparza [10], Verma *et al.* [32] has given a linear translation from context-free grammars (equivalently, pushdown automata) to the NP-complete existential fragment of Presburger arithmetic expressing their Parikh images. Such a translation has been used to derive optimal algorithms for equational horn clauses [32]. This linear translation to existential Presburger formulas is orthogonal to our result. On the one hand, our algorithm works only on NFAs and has exponential time complexity when the alphabet size is not fixed and therefore does not immediately yield an NP algorithm. On the other hand, existential Presburger formulas are exponentially more succinct than the classical representation of semilinear sets (as a union of linear sets) and it is easy to check that the translation in [32] does not produce output formulas in any known polynomial-time fragment of existential Presburger arithmetic even when the input is an NFA (or its equivalent CFG) over some fixed alphabet size. We will simply remark that many scenarios (including [13, 14, 16, 20, 26, 30] and the ones we consider in this paper) require the use of the classical representation of semilinear sets to derive optimal algorithms.

*Vector generalizations* of the Frobenius problems (e.g. see [25] and references therein) have also been proposed. Such generalizations are orthogonal to our results for semilinear sets. For one, they mainly attempt to extend the notions of "conductors" from the original Frobenius problem, but neither attempt to reduce the number of generators nor obtain an efficient algorithm for doing so.

**Organization** We fix our notations in Section 2. We prove a normal form theorem for semilinear sets in Section 3, which is applied in Section 4 to prove a general version of Chrobak-Martinez Theorem. Lower bounds are given in Section 5, and applications are given in Section 6.

**Acknowledgment** The author thanks Shunichi Amano, Floris Geerts, Stefan Göller, Filip Murlak, Benjamin Rubinstein, and Tony Tan for their helpful comments. The author is supported by EPSRC grant E005039.

## 2 Preliminary

**General notations** Let $\mathbb{N}$ denote the set of nonnegative integers. We assume familiarity with basic notions from linear algebra: vector space, basis, linear independence, rank, etc. In the sequel, we use only the standard real vector space $\mathbb{R}^k$. We use $\mathbf{0}$ to denote the vector of $\mathbb{R}^k$ with all-zero entries. We denote by $\{\mathbf{e}_i\}_{i=1}^k$ the standard basis for $\mathbb{R}^k$, where $\mathbf{e}_i$ denotes the vector with all-zero entries except for the $i$th. We shall also require one notation from the theory of convex sets [33]. Given a finite set $V = \{\mathbf{v}_1, \ldots, \mathbf{v}_m\}$ of vectors in $\mathbb{R}^k$, let $\mathbf{cone}(V)$ be the set $\{\Sigma_{i=1}^m \lambda_i \mathbf{v}_i : \lambda_i \in \mathbb{R}_{\geq 0}\}$.

**Partial orders** Recall that a partial order $\preceq$ on a set $S$ is *well-founded* if there does not exist a strictly decreasing infinite sequence $s_1 \succ s_2 \succ \ldots$ of elements from $S$. An element $s$ of $S$ is said to be $\preceq$-*minimal*, if all $s' \in S$ with $s' \preceq s$ satisfies $s = s'$.

In the sequel, we shall reserve $\preceq$ for the component-wise partial order on $\mathbb{N}^k$, i.e., $(a_1, \ldots, a_k) \preceq (b_1, \ldots, b_k)$ iff $a_i \leq b_i$ for all $i \in \{1, \ldots, k\}$. Dickson's lemma [8] states that $\preceq$ is well-founded.

**Automata** An alphabet $\Sigma$ is simply a finite set of *letters*. An NFA $A$ is a tuple $(\Sigma, Q, \delta, q_0, q_F)$, where $Q$ is a finite set of states, $q_0 \in Q$ is an initial state, $q_F \in Q$ is the final state, and $\delta \subseteq Q \times \Sigma \times Q$ is a transition relation. A *path* $\pi$ in $A$ from $q \in Q$ to $q' \in Q$ is simply an alternating sequence $p_0 \beta_1 p_1 \ldots \beta_m p_m \in (Q\Sigma)^* Q$ of states and letters such that $p_0 = q$ and $p_m = q'$. In this case, we write $L(\pi)$ to denote the *path labels* $\beta_1 \ldots \beta_m \in \Sigma^*$, and say that $\pi$ is a *path on* $\beta_1 \ldots \beta_m$. For convenience, we shall sometimes omit the path labels from $\pi$, and simply refer to it as a path $\pi = p_0 \ldots p_m$ on word $w = \beta_1 \ldots \beta_m$. In this case, the length $|\pi|$ (resp. $|w|$) of the path $\pi$ (resp. word $w$) is $m$. Given $i, j \in \{0, \ldots, |\pi|\}$ with $i \leq j$, we also write $\pi[i,j]$ as the *path segment* $p_i p_{i+1} \ldots p_j$ of $\pi$. Given two paths $\pi = p_0 \beta_1 \ldots p_m$ and $\pi' = p_m \beta_{m+1} \ldots p_n$ (with $m \leq n$), we let $\pi \odot \pi'$ denote the concatenated path $p_0 \beta_1 \ldots p_m \beta_{m+1} \ldots p_n$. The NFA $A$ is said to *accept* the word $w \in \Sigma^*$ if there exists a path $\pi$ in $A$ from $q_0$ to $q_F$ such that $L(\pi) = w$. In this case, $\pi$ is said to be an *accepting path*. The *language* $L(A)$ of $A$ is the set of words accepted by $A$.

Given a word $w \in \Sigma^*$ and $\alpha \in \Sigma$, we write $|w|_\alpha$ to denote the number of occurences of $\alpha$ in $w$. In the sequel, we tacitly assume that there is a linear ordering $\leq$ on $\Sigma$, i.e., $\Sigma = \{\alpha_1, \ldots, \alpha_k\}$ with $\alpha_i \leq \alpha_j$ iff $i \leq j$. In this case, given a word $w \in \Sigma^*$, we define the *Parikh image* $\mathcal{P}(w)$ of $w$ to be the tuple $(|w|_{\alpha_1}, \ldots, |w|_{\alpha_k}) \in \mathbb{N}^k$. In addition, given a set $L \subseteq \Sigma^*$, we define the Parikh image $\mathcal{P}(L)$ of $L$ to be the set $\{\mathcal{P}(w) : w \in L\} \subseteq \mathbb{N}^k$. We shall also write $\mathcal{P}(\pi)$ for a path $\pi$ in an NFA to denote $\mathcal{P}(L(\pi))$.

**Matrix notations** Given two $n$-by-$n$ 0-1 matrices $M = [m_{i,j}]_{n \times n}$ and $M' = [m'_{i,j}]_{n \times n}$, we write $M \bullet M'$ to denote the matrix $M'' = [m''_{i,j}]_{n \times n}$ with $m''_{i,j} = \bigvee_{k=1}^n (m_{i,k} \wedge m'_{k,j})$. The operator $\bullet$ is often referred to as *boolean matrix multiplication*, which can easily be evaluated in $O(n^3)$. We also write $M \vee M'$ to denote the application of the boolean operation $\vee$ component-wise, i.e., resulting in a matrix $M'' = [m''_{i,j}]_{n \times n}$ with $m''_{i,j} = m_{i,j} \vee m'_{i,j}$. In the sequel, we shall also write $M[i,j]$ for the $(i,j)$-component $m_{i,j}$ of $M$.

**Semilinear sets** For every vector $\mathbf{v} \in \mathbb{Z}^k$ and every finite set $S = \{\mathbf{u}_1, \ldots, \mathbf{u}_m\}$ of vectors in $\mathbb{Z}^k$, we write $P(\mathbf{v}; S)$ to denote the $\mathbb{Z}$-*linear set* $\{\mathbf{v} + \Sigma_{i=1}^m a_i \mathbf{u}_i : a_1, \ldots, a_m \in \mathbb{N}\}$. The pair $B := \langle \mathbf{v}; S \rangle$ is said to be a *linear basis* for $P(\mathbf{v}; S)$. Notice that there exist non-unique linear bases for a $\mathbb{Z}$-linear set. The vector $\mathbf{v}$ is said to be the *offset* of $B$, and the vectors $S$ the *generators* of $B$. For $\mathbf{v} = \mathbf{0}$, we also use $\mathbf{cone}_\mathbb{N}(S)$ to denote $P(\mathbf{v}; S)$. A $\mathbb{Z}$-semilinear set $S$ is simply a finite (possibly empty) union of $\mathbb{Z}$-linear sets $P(\mathbf{v}_1; S_1), \ldots, P(\mathbf{v}_s; S_s)$. In this case, we say that $\mathcal{B} = \{\langle \mathbf{v}_i; S_i \rangle\}_{i=1}^s$ is a *semilinear basis* for $P(\mathcal{B}) := S$. Likewise, semilinear bases for $S$ are not unique. A $\mathbb{Z}$-semilinear set $S \subseteq \mathbb{Z}^k$ is said to be $\mathbb{N}$-*semilinear* (or simply *semilinear*) if it has a semilinear basis with vectors from $\mathbb{N}^k$ only. The notion of $\mathbb{N}$-*linear* (or simply *linear*) sets is also defined similarly.

Since $\mathbb{Z}$-(semi)linear bases $\mathcal{B}$ are simply a sequence of vectors from $\mathbb{Z}^k$, we could talk about their size $\|\mathcal{B}\|$ when represented on the tapes of Turing machines. We shall use both *unary* and *binary* representations of numbers, and be explicit about this when necessary. In the sequel, we shall *not* distinguish (semi)linear sets and their bases, when it is clear from the context. Thus, we shall use such a phrase as "compute a (semi)linear set" to mean that we compute a particular (semi)linear basis for it.

**Arithmetic on $2^{\mathbb{Z}^k}$** Given two sets $S_1, S_2 \subseteq \mathbb{Z}^k$, we define an operation '+' on them as follows: $S_1 + S_2 := \{\mathbf{v}_1 + \mathbf{v}_2 : \mathbf{v}_1 \in S_1, \mathbf{v}_2 \in S_2\}$. Suppose now that $S_1 = \bigcup_{i=1}^r P(\mathbf{v}_i; V_i)$ and $S_2 = \bigcup_{j=1}^t P(\mathbf{w}_j; W_j)$. Then, observe that $S_1 + S_2 = \bigcup_{i=1}^r \bigcup_{j=1}^t P(\mathbf{v}_i + \mathbf{w}_j; V_i \cup W_j)$. For $s_1 \in S_1$, we shall also write $s_1 + S_2$ to mean $\{s_1\} + S_2$. Thus, we have $P(\mathbf{v}; S) = \mathbf{v} + \mathbf{cone}_\mathbb{N}(S)$.

# 3 Normal form for semilinear sets

In this section, we shall prove a normal form theorem for semilinear sets: given a semilinear basis $\mathcal{B}$ in $\mathbb{N}^k$ represented in unary, we can compute in time polynomial in $\|\mathcal{B}\|$ and exponential in $k$ another semilinear basis $\mathcal{B}' = \{\langle \mathbf{w}_i; S_i \rangle\}_{i=1}^m$ in unary such that $P(\mathcal{B}) = P(\mathcal{B}')$ and $|S_i| \leq k$ for each $i \in \{1, \ldots, m\}$. For simplicity, we shall state only a general theorem for $\mathbb{Z}$-linear sets of the form $\mathbf{cone}_{\mathbb{N}}(V)$, where $V$ is a finite subset of $\mathbb{Z}^k$; this can easily be used to derive the desired normal form theorem for semilinear sets (among others). [Recall that $P(\mathbf{v}; V) = \mathbf{v} + \mathbf{cone}_{\mathbb{N}}(V)$.]

**Theorem 3.1** *Let $V := \{\mathbf{v}_1, \ldots, \mathbf{v}_m\} \subseteq \mathbb{Z}^k \setminus \{\mathbf{0}\}$ with $m > 0$. Let $a \in \mathbb{N}$ be the maximum absolute value of numbers appearing in vectors of $V$. Then, it is possible to compute in time $2^{O(k \log(m) + k^2 \log(ka))}$ a sequence of $\mathbb{Z}$-linear bases $\langle \mathbf{w}_1; S_1 \rangle, \ldots, \langle \mathbf{w}_s; S_s \rangle$ such that*

$$\mathbf{cone}_{\mathbb{N}}(V) = \bigcup_{i=1}^s P(\mathbf{w}_i; S_i)$$

*where the maximum absolute value of entries of each $\mathbf{w}_i$ is $O(m(k^2 a)^{2k+3})$, each $S_i$ is a subset of $V$ with $|S_i| \leq k$, and $s = O(m^{2k}(k^2 a)^{2k^2+3k})$. Furthermore, if $V \subseteq \mathbb{N}^k$, we have $\{\mathbf{w}_1, \ldots, \mathbf{w}_s\} \subseteq \mathbb{N}^k$.*

Observe that this theorem causes only an exponential blow-up in the dimension $k$. Moreover, each set $S_i$ contains at most $k$ generators. To prove this theorem, we start with a slight strengthening of *the conical version of Caratheodory's theorem* from the theory of convex sets [33, Proposition 1.15]. Proof is given in the appendix.

**Lemma 3.2** *Let $V := \{\mathbf{v}_1, \ldots, \mathbf{v}_m\} \subseteq \mathbb{Z}^k \setminus \{\mathbf{0}\}$ with $m > 0$. Let $a \in \mathbb{N}$ be the maximum absolute value of numbers appearing in vectors of $V$. Then, it is possible to compute in time $2^{O(k \log m + \log a)}$, a sequence $S_1, \ldots, S_r$ of distinct linearly independent $d$-subsets of $V$, where $d \in \{1, \ldots, k\}$ is the rank of $V$, and*

$$\mathbf{cone}(V) = \bigcup_{i=1}^r \mathbf{cone}(S_i).$$

Let us first explain the idea behind the rest of the proof of Theorem 3.1. Intuitively, Lemma 3.2 says that $\mathbf{cone}(V) \subseteq \mathbb{R}^k$ can be subdivided into smaller subcones with exactly $d \in \{1, \ldots, k\}$ generators where $d$ is the rank of $V$. This lemma immediately gives an *upper bound* for $\mathbf{cone}_{\mathbb{N}}(V)$ as the union of the *integer points* in $\mathbf{cone}(S_i)$; in general, the latter contains much more points than $\mathbf{cone}_{\mathbb{N}}(V)$. On the other hand, we have $\bigcup_{i=1}^r \mathbf{cone}_{\mathbb{N}}(S_i) \subseteq \mathbf{cone}_{\mathbb{N}}(V)$, where the inclusion is strict in general. It turns out that an equality can be achieved by first making a "few" *duplicates* of each $\mathbf{cone}_{\mathbb{N}}(S_i)$ and then *shifting* them appropriately by some "small" integer vectors.

We now prove Theorem 3.1. First invoke Lemma 3.2 on $V$ and obtain linearly independent $d$-subsets $S_1, \ldots, S_r$ of $V$, where $d = \mathbf{rank}(V)$ and $r \leq m^k$, satisfying $\mathbf{cone}(V) = \bigcup_{j=1}^r \mathbf{cone}(S_j)$. Then, it follows that $\mathbf{cone}(V) \cap \mathbb{Z}^k = \bigcup_{j=1}^r \left( \mathbf{cone}(S_j) \cap \mathbb{Z}^k \right)$. To compute the integer vector "shifts", we shall need to define the notions of *canonical* and *minimal* vectors.

**Characterization via canonical and minimal vectors**

Suppose now that $\mathbf{v} \in \mathbf{cone}(S_j) \cap \mathbb{Z}^k$ and $S_j = \{\mathbf{u}_1, \ldots, \mathbf{u}_d\}$. We make several simple observations:

- **(O1)** There exists a *unique* vector $[\mathbf{v}] \in \{-ka, \ldots, ka\}^k \cap \mathbf{cone}(S_j)$ and *unique* non-negative integers $a_1, \ldots, a_d$ such that: 1) $\mathbf{v} = [\mathbf{v}] + \Sigma_{i=1}^d a_i \mathbf{u}_i$, and 2) $[\mathbf{v}] = \Sigma_{i=1}^d b_i \mathbf{u}_i$ for some (unique) $0 \leq b_1, \ldots, b_d < 1$. To see this, observe that by linear independence of $S_j$ there exist some unique $\lambda_1, \ldots, \lambda_d \in \mathbb{R}_{\geq 0}$ such that $\mathbf{v} = \Sigma_{i=1}^d \lambda_i \mathbf{u}_i$. Simply let $a_i := \lfloor \lambda_i \rfloor$, $b_i := \lambda_i - a_i$, and $[\mathbf{v}] := \Sigma_{i=1}^d b_i \mathbf{u}_i$. Uniqueness is immediate from uniqueness of $\lambda_1, \ldots, \lambda_d$.

- **(O2)** Given $\mathbf{v}' \in \mathbf{cone}(S_j) \cap \mathbb{Z}^k$, we write $\mathbf{v} \sim \mathbf{v}'$ iff $[\mathbf{v}] = [\mathbf{v}']$. It is easy to see that $\sim$ is an equivalence relation of finite index (there are at most $(2ka+1)^k$ equivalence classes). If $[\mathbf{v}] = \mathbf{v}$, the vector $\mathbf{v}$ is said to be a *canonical representative* of the equivalence class $\{\mathbf{u} \in \mathbf{cone}(S_j) \cap \mathbb{Z}^k : [\mathbf{u}] = \mathbf{v}\}$. In this case, we will also call $\mathbf{v}$ an $S_j$-*canonical vector*, or simply *canonical vector* when $S_j$ is understood.

- **(O3)** If $\mathbf{v}$ is in $\mathbf{cone}_{\mathbb{N}}(V) \cap \mathbf{cone}(S_j)$, then $\mathbf{v} + \Sigma_{i=1}^d c_i \mathbf{u}_i \in \mathbf{cone}_{\mathbb{N}}(V) \cap \mathbf{cone}(S_j)$ for every $c_1, \ldots, c_d \in \mathbb{N}$.

We shall now use these observations to define a natural well-founded partial order $\trianglelefteq_j$ on $\mathbf{cone}_{\mathbb{N}}(V) \cap \mathbf{cone}(S_j)$; note that $\mathbf{cone}_{\mathbb{N}}(V) \cap \mathbf{cone}(S_j) \neq \emptyset$. Given $\mathbf{v}, \mathbf{w} \in \mathbf{cone}_{\mathbb{N}}(V) \cap \mathbf{cone}(S_j)$, we write $\mathbf{v} \trianglelefteq_j \mathbf{w}$ iff, for some (unique) $S_j$-canonical vector $\mathbf{v}_0$ and some (unique) coefficients $a_1, \ldots, a_d \in \mathbb{N}$ and $b_1, \ldots, b_d \in \mathbb{N}$, it is the case that: 1) $\mathbf{v} = \mathbf{v}_0 + \Sigma_{i=1}^d a_i \mathbf{u}_i$, 2) $\mathbf{w} = \mathbf{v}_0 + \Sigma_{i=1}^d b_i \mathbf{u}_i$, and 3) $(a_1, \ldots, a_d) \preceq (b_1, \ldots, b_d)$. The following simple lemma (proof in the appendix) shows that $\trianglelefteq_j$ is a well-founded partial order, and characterizes $\trianglelefteq_j$-minimal elements.

**Lemma 3.3** *The relation $\trianglelefteq_j$ is a well-founded partial order on $\mathbf{cone}_{\mathbb{N}}(V) \cap \mathbf{cone}(S_j)$. Furthermore, a vector $\mathbf{v} \in \mathbf{cone}_{\mathbb{N}}(V) \cap \mathbf{cone}(S_j)$ is $\trianglelefteq_j$-minimal iff none of the vectors $(\mathbf{v} - \mathbf{u}_1), \ldots, (\mathbf{v} - \mathbf{u}_d)$ are in $\mathbf{cone}_{\mathbb{N}}(V) \cap \mathbf{cone}(S_j)$.*

Lemma 3.3 and Observation **(O3)** immediately implies that $\mathbf{cone}_{\mathbb{N}}(V)$ is a union of linear sets $P(\mathbf{v}; S_j)$ taken over all $j = 1, \ldots, r$ and $\unlhd_j$-minimal vectors $\mathbf{v}$.

**Lemma 3.4** $\mathbf{cone}_{\mathbb{N}}(V) = \bigcup_{j=1}^{r} \bigcup_{\mathbf{v}} P(\mathbf{v}; S_j)$, *where* $\mathbf{v}$ *is taken over all* $\unlhd_j$-*minimal vectors.*

**Proof.** ($\supseteq$) Obvious.

($\subseteq$) If $\mathbf{v} \in \mathbf{cone}_{\mathbb{N}}(V)$, then $\mathbf{v} \in \mathbf{cone}(S_j) \cap \mathbb{Z}^k$ for some $j \in \{1, \ldots, r\}$. By Lemma 3.3, there exists a $\unlhd_j$-minimal vector $\mathbf{v}'$ satisfying $\mathbf{v}' \unlhd_j \mathbf{v}$. Observation **(O3)** implies that $\mathbf{v} \in P(\mathbf{v}'; S_j)$. □

Note also that if $V \subseteq \mathbb{N}^k$, then all $\unlhd_j$-minimal vectors ($1 \leq j \leq r$) are also nonnegative.

A roadmap for rest of the proof is as follows. We shall show that each $\unlhd_j$-minimal vectors cannot be too large and can be efficiently enumerated. This will immediately give us the desired sequence of linear bases. The proof of this will require connections to integer programming, and the use of dynamic programming.

**Bounds via integer programming**

For each $S_j = \{\mathbf{u}_1, \ldots, \mathbf{u}_d\}$, we shall now show that all $\unlhd_j$-minimal vectors $\mathbf{v}$ cannot be too large. To this end, for each canonical vector $\mathbf{v}_0 \in \mathbf{cone}(S_j) \cap \mathbb{Z}^k$, consider the integer linear program $A\mathbf{x} = \mathbf{v}_0^{\mathrm{T}}$ ($\mathbf{x} \succeq \mathbf{0}$), where $A$ is the $k \times (m+d)$ matrix consisting of columns $\mathbf{v}_1, \mathbf{v}_2, \ldots, \mathbf{v}_m, -\mathbf{u}_1, -\mathbf{u}_2, \ldots, -\mathbf{u}_d$ (in this order) and $\mathbf{x}$ is the column $(m+d)$-vector consisting of the variables $x_1, \ldots, x_m, y_1, \ldots, y_d$ (in this order). The following simple lemma (whose proof is in the appendix) shows that $\preceq$-minimal solutions to such integer programs — as we shall see, they cannot be too large as well — provide upper bounds for how large $\unlhd_j$-minimal vectors can be.

**Lemma 3.5** *For every* $\unlhd_j$-*minimal vector* $\mathbf{v} \in \mathbf{cone}_{\mathbb{N}}(V) \cap \mathbf{cone}(S_j)$, *let* $\mathbf{v}_0 := [\mathbf{v}]$ *and* $\mathbf{b} = (b_1, \ldots, b_d) \in \mathbb{N}^d$ *be the unique coefficients such that* $\mathbf{v} = \mathbf{v}_0 + \Sigma_{i=1}^k b_i \mathbf{u}_i$. *Suppose also that* $\mathbf{c} = (c_1, \ldots, c_m)$ *is a* $\preceq$-*minimal solution to the integer program* $\Sigma_{i=1}^m x_i \mathbf{v}_i = \mathbf{v}$ ($\mathbf{x} \succeq \mathbf{0}$)*. Then, the vector* $\mathbf{w} := (\mathbf{c}, \mathbf{b}) \in \mathbb{N}^{m+d}$ *is a* $\preceq$-*minimal solution to the integer program* $A\mathbf{x} = \mathbf{v}_0$ ($\mathbf{x} \succeq \mathbf{0}$)*.*

Consider the set $U$ of all vectors $\mathbf{v}_0 + \Sigma_{i=1}^d a_i \mathbf{u}_i$, where $\mathbf{v}_0$ ranges over all canonical vectors and $a_1, \ldots, a_d$ ranges over all $d$-tuples of nonnegative integers such that $(c_1, \ldots, c_m, a_1, \ldots, a_d)$ is a $\preceq$-minimal solution to the integer program $A\mathbf{x} = \mathbf{v}_0$, for some $c_1, \ldots, c_m \in \mathbb{N}$. We shall see now that the maximum absolute value $B$ of numbers appearing in $U$ exist, which immediately gives an upper bound for the maximum absolute value of entries of $\unlhd_j$-minimal vectors. The following general lemma, whose proof is a straightforward adaptation of the proof of [21, Theorem p. 767], yields an upper bound for $B$.

**Lemma 3.6** *Let* $A$ *be a* $k \times n$ *integer matrix and* $\mathbf{b}$ *a* $k$-*vector, both with entries in* $[-t, t] \cap \mathbb{Z}$, *where* $t \in \mathbb{N}$. *Then, every* $\preceq$-*minimal solution* $\mathbf{x} \in \mathbb{N}^n$ *to* $A\mathbf{x} = \mathbf{b}$ ($\mathbf{x} \succeq \mathbf{0}$) *is in* $\{0, 1, \ldots, n(kt)^{2k+1}\}^n$.

Notice that the maximum absolute value of numbers appearing in our integer programs cannot exceed $t := ak$ (which could appear on the right hand side of the equation). If $M := (m+k)(kt)^{2k+1}$, it follows that $B \leq akM + ak \leq N := (m+k)(k^2a)^{2k+2} + ak$. This completes the proof of *existence* for Theorem 3.1 and gives us the desired bounds for the parameter $s$ and the maximum absolute value of entries of each $w_i$ in Theorem 3.1. It remains to show how to make this algorithmic.

**Computing canonical and minimal vectors**

We first show how to compute all the canonical vectors. Since Gaussian-elimination over rational numbers can be implemented to run in time polynomial in the total number of bits in the input matrix [9] and that each $S_j$ is linearly independent, we could easily compute all $S_j$-canonical vectors (for all $j \in \{1, \ldots, r\}$) in time $2^{O(k \log(ka) + k \log m)}$ by going through all candidate vectors $\mathbf{v} \in \{-ka, \ldots, ka\}^k$ and checking whether there exist $0 \leq b_1, \ldots, b_d < 1$ such that $\Sigma_{i=1}^d b_i \mathbf{u}_i = \mathbf{v}$. [Transform into row-reduced echelon form to compute the *unique* solution, if exists. Since $S_j \cup \{\mathbf{v}\} \subseteq \mathbb{Z}^k$, the coefficients $b_1, \ldots, b_d$ will be rational.]

For each fixed $j \in \{1, \ldots, r\}$ and each fixed $S_j$-canonical vector $\mathbf{v}_0$, we now show how to compute the set of all $\unlhd_j$-minimal vectors $\mathbf{v}$ such that $[\mathbf{v}] = \mathbf{v}_0$ by dynamic programming in time $2^{O(k \log m + k^2 \log(ka))}$. Observe that since there are at most $r(2ak+1)^k = 2^{O(k \log(kam))}$ possible $\mathbf{v}_0$, doing this for *all* canonical vectors would take time $2^{O(k \log m + k^2 \log(ka))}$, which is also the total complexity of the algorithm. To this end, we first fill out *in stages* a table $T_1$ which keeps track of all vectors $\mathbf{v} \in \{0, 1, \ldots, N\}^k \cap \mathbf{cone}_{\mathbb{N}}(V)$. At stage $h = 1, 2, \ldots, m$, we collect all vectors $\mathbf{v}$ that can be written as $\Sigma_{i=1}^h c_i \mathbf{v}_i$, where $0 \leq c_i \leq M$. Since the size of the table is at most $N^k (k \log N)$ — $k \log N$ bits are used to identify each element in the table with an associated $k$-tuple — this could be carried out in time $O(m(N^k(k \log N))^2) = 2^{O(k \log m + k^2 \log(ka))}$. We then fill out in stages another table $T_2$, which keeps track of all vectors $\mathbf{v} \in \{0, 1, \ldots, N\}^k \cap P(\mathbf{v}_0; S_j)$. This could be done in $d$ stages, similar to the computation of $T_1$, and could be implemented to run in time $O(k(N^k(k \log N))^2) = 2^{O(k \log m + k^2 \log(ka))}$. We then simply compute a new table $T_3 = T_1 \cap T_2$, from which we eliminate vectors that are not $\unlhd_j$-minimal by using the characterization of $\unlhd_j$-minimal vectors from Lemma 3.3. All in all, this could be implemented to run in time $2^{O(k \log m + k^2 \log(ka))}$.

## 4 Parikh Images of NFAs

In this section, we shall apply Theorem 3.1 to obtain the main result for Parikh images of NFAs.

**Theorem 4.1** *Let $A$ be an NFA with $n$ states over an alphabet $\Sigma$ of size $k$. Then, there exists a representation of $\mathcal{P}(A)$ as a union of linear sets $P(\mathbf{v}_1; S_1), \ldots, P(\mathbf{v}_m; S_m)$, where the maximum entry of each $\mathbf{v}_i$ is $O(n^{3k+5} k^{4k+6})$, each $S_i$ is a subset of $\{0, \ldots, n\}^k$ with $|S_i| \leq k$, and $m = O(n^{k^2+3k+5} k^{4k+6})$. Furthermore, this is computable in time $2^{O(k^2 \log(kn))}$.*

Observe that this theorem causes an exponential blow-up only in the size of the alphabet. Efficiency could be improved by outputting numbers in binary.

We shall devote the rest of this section to prove this theorem. Let $A = (\Sigma, Q, \delta, q_0, q_F)$ be a given NFA, where $|Q| = n$ and $\Sigma = \{\alpha_1, \ldots, \alpha_k\}$. Throughout the proof, we shall use the notion of "cycle type". A *cycle type* is a Parikh image $\mathbf{v} \in \mathbb{N}^k$ of any word $w \in \Sigma^{\leq n}$ such that there is a path $\pi$ of $A$ on $w$ from some (not necessarily initial) state $p$ to itself. The cycle $\pi$ is said to *witness* $\mathbf{v}$. Observe that the sum of the components of any cycle type cannot exceed $n$.

**Characterization of $\mathcal{P}(L(A))$**

We shall first give a characterization of the Parikh image of $A$ in terms of Parikh images of "short" paths together with some cycle types. Given a path $\pi = p_0 \beta_1 p_1 \ldots \beta_r p_r$ of $A$ from the (not necessarily initial) state $p_0$ to the state $p_r$, let $S_\pi = \subseteq \{0, \ldots, n\}^k$ be the set of all the cycle types that are witnessed by some cycles $C = p'_0 p'_1 \ldots p'_t p_0$ in $A$ such that $p'_i = p_j$ for some $i \in \{0, \ldots, t\}$ and $j \in \{0, \ldots, r\}$. That is, $C$ and $\pi$ *meet* at state $p'_i = p_j$. Now define $T_\pi$ to be the linear set $P(\mathcal{P}(\pi); S_\pi)$.

**Lemma 4.2** *The following identity holds:*
$$\mathcal{P}(L(A)) = \bigcup_\pi T_\pi,$$
*where $\pi$ is taken over all accepting path of $A$ of length at most $(n-1)^2$.*

**Proof.** ($\subseteq$) Assume that $\mathbf{v} \in \mathcal{P}(L(A))$ and $\sigma = p_0 \beta_1 p_1 \ldots \beta_r p_r$ be an accepting path in $A$ such that $\mathcal{P}(\sigma) = \mathbf{v}$. We shall construct another accepting path $\sigma'$ in $A$ of length at most $(n-1)^2$. For each state $q$ occurring in $\sigma$, let $l(q)$ be the last (i.e. maximum) index $i \in \{0, \ldots, r\}$ such that $p_i = q$. Let us write down all such $l(q)$ in an increasing order, e.g., $i_0 < i_1 < \ldots < i_s = r$. Note that $s < n$. By a standard result in graph theory, each subpath $\sigma[i_j, i_{j+1}]$ of $\sigma$ can be decomposed into a simple path $\pi_j$ from $p_{i_j}$ to $p_{i_{j+1}}$ of length at most $n-1$ and finitely many simple cycles (possibly with duplicates) $C_1, \ldots, C_h$ each of length at most $n$. That is, we have $\mathcal{P}(\pi[i_j, i_{j+1}]) = \mathcal{P}(\pi_j) + \Sigma_{i=1}^h \mathcal{P}(C_i)$. Such a decomposition result, however, might allow some cycle $C_i$ to *avoid* (i.e. not meet with) $\pi_j$ as can be witnessed from [29, Figure 2]. On the other hand, $C_i$ must visit some states of $p_{i_0}, \ldots, p_{i_s}$ as these contain all states in $\sigma$. Thus, we simply define $\sigma'$ to be the accepting path $\pi_0 \odot \pi_1 \odot \ldots \odot \pi_{s-1}$ of length at most $(n-1)^2$. It follows that $\mathbf{v} \in T_{\sigma'}$.

($\supseteq$) Conversely, let $\mathbf{v} \in T_\pi$ for some accepting path $\pi$ in $A$ of length at most $(n-1)^2$. Then, if $S_\pi = \{\mathbf{v}_1, \ldots, \mathbf{v}_s\}$, then $\mathbf{v} = \mathcal{P}(\pi) + \Sigma_{i=1}^s a_i \mathbf{v}_i$ for some $a_1, \ldots, a_s \in \mathbb{N}$. Let $C_i$ be a cycle in $A$ that meets with $\pi$ and satisfies $\mathcal{P}(C_i) = \mathbf{v}_i$. We can construct an accepting path $\sigma$ in $A$ with $\mathcal{P}(\sigma) = \mathbf{v}$ as follows: start from $\pi$ as the "base" path, and for each $i \in \{1, \ldots, s\}$, attach $a_i$ copies of $C_i$ to one pre-selected common state of $C_i$ and $\pi$. □

As an immediate corollary, we have:

**Proposition 4.3** *Let $A$ be an NFA with $n$ states over an alphabet $\Sigma$ of size $k$. Then, $\mathcal{P}(A)$ can be represented as a union of linear sets $P(\mathbf{v}_1; S_1), \ldots, P(\mathbf{v}_m; S_m)$, where $\mathbf{v}_i \in \{0, \ldots, (n-1)^2\}^k$ and the components of each vector in $S_i$ cannot exceed $n$.*

*Remark:* A slightly stronger version of this proposition was claimed in [26], where the maximum component of each $\mathbf{v}_i$ cannot exceed $n$. Their proof turns out to have a subtle error that also occurs in the proof of Chrobak-Martinez Theorem [6, 20], which was recently fixed in [29]. In fact, we show in Proposition 5.3 that our quadratic bound is essentially optimal, i.e., it cannot be lowered to $o(n^2)$. (*End Remark*)

Observe now that the proof of existence in Theorem 4.1 is essentially immediate from Proposition 4.3 and Theorem 3.1. In fact, if we are only interested in existence, this yields a better upper bound of $O(n^{3(k+1)} k^{4k+6})$ (resp. $O(n^{k^2+3k+3} k^{4k+6})$) for the maximum entry of all offsets $\{\mathbf{v}_i\}_{i=1}^m$ (resp. $m$) in the statement of Theorem 4.1.

**Dynamic programming algorithm**

It remains to show that this can be made algorithmic. To this end, we first show how to compute all the cycle types of $A$. More precisely, let $Q = \{q_0, \ldots, q_{n-1}\}$, where $q_{n-1} := q_F$, and let $I = \{(b_1, \ldots, b_k) \in \mathbb{N}^k : \Sigma_{i=1}^k b_i \leq n\}$. For each vector $\mathbf{v} \in \mathbb{N}^k$, we write $M_\mathbf{v} = [a_{i,j}]_{n \times n}$ for the $n$-by-$n$ 0-1 matrix where $a_{i,j} = 1$ iff there exists a path $\pi$ from $q_i$ to $q_j$ with $\mathcal{P}(\pi) = \mathbf{v}$. We are interested in computing all matrices $M_\mathbf{v}$ for each $\mathbf{v} \in I$. Observe that the naive algorithm, which runs through all paths of $A$ from $q_i$ to $q_j$ with $\mathcal{P}(\pi) = \mathbf{v}$, has time complexity that is exponential in $n$. We will give an algorithm for computing these in time $2^{O(k \log n)}$ using dynamic programming. To this end, let us derive a recursive formula for computing $M_{\mathbf{v}}$ based on $M_{\mathbf{v}'}$ with $\mathbf{v}' \preceq \mathbf{v}$. As base cases, we first observe that $M_{\mathbf{0}}$ is the $n$-by-$n$ identity matrix. Furthermore, each matrix $M_{\mathbf{e}_i}$ could be constructed easily from the transition relation $\delta$.

**Lemma 4.4** *Let* $\mathbf{v} = (r_1, r_2, \ldots, r_{i-1}, r_i+1, r_{i+1}, \ldots, r_k)$ *with each* $r_i \in \mathbb{N}$*. Then, the following identity holds:*

$$M_{\mathbf{v}} = \bigvee_{\mathbf{u}, \mathbf{w}} M_{\mathbf{u}} \bullet M_{\mathbf{e}_i} \bullet M_{\mathbf{w}}$$

*where* $\mathbf{u}$ *ranges over all vectors* $\preceq \mathbf{v}$ *whose ith entry is 0, and* $\mathbf{w}$ *is the vector* $\mathbf{v} - \mathbf{e}_i - \mathbf{u}$*.*

The proof of this lemma can be found in the appendix. Intuitively, this recurrence relation can be derived by observing that a path $\pi$ with $\mathcal{P}(\pi) = \mathbf{v}$ can be *uniquely* decomposed into three consecutive path segments $\pi_1$, $\pi_2$, and $\pi_3$ with $\mathcal{P}(\pi_1) = \mathbf{u}$, $\mathcal{P}(\pi_2) = \mathbf{e}_i$, and $\mathcal{P}(\pi_3) = \mathbf{w}$, for some prescribed $\mathbf{u}$ and $\mathbf{w}$. The following lemma, whose proof is also in the appendix, is a simple application of Lemma 4.4 and dynamic programming.

**Lemma 4.5** *We can compute* $\{M_{\mathbf{v}}\}_{\mathbf{v} \in I}$ *in time* $2^{O(k \log n)}$*.*

For each $i \in \{0, \ldots, n-1\}$, let $\Gamma_i$ be the set of all cycle types $\mathbf{v}$ witnessed by some cycle $\pi = p_0 p_1 \ldots p_0$ in $A$ with $p_j = q_i$. Since $\{M_{\mathbf{v}}\}_{\mathbf{v} \in I}$ have been computed, all sets $\Gamma_i$ could be computed within $O(n^{k+1})$ extra time.

We now show how to compute $\mathcal{P}(L(A))$ in time $2^{O(k^2 \log(kn))}$. To this end, we shall use another application of dynamic programming based on Lemma 4.2, Theorem 3.1, and the sets $\{\Gamma_i\}_{i=0}^{n-1}$, which we already computed. For each $0 \leq i \leq (n-1)^2$ and each $0 \leq j < n$, let $T_{i,j} := \bigcup_{\pi} T_{\pi}$ where $\pi$ is taken over all paths in $A$ of length $i$ from $q_0$ to $q_j$. By Lemma 4.2, it is the case that $\mathcal{P}(L(A)) = \bigcup_{i=0}^{(n-1)^2} T_{i,n-1}$; recall that $q_{n-1} = q_F$ by definition. We shall now derive a recurrence relation for $T_{i,j}$.

**Lemma 4.6** *It is the case that* $T_{0,0} = \{\mathbf{0}\}$ *and* $T_{0,j} = \emptyset$ *for each* $j \in \{1, \ldots, n-1\}$*. Whenever* $i > 0$ *and* $j \in \{0, \ldots, n-1\}$*, we have*

$$T_{i,j} = \bigcup_{h=0}^{n-1} \left( T_{i-1,h} + \bigcup_{1 \leq l \leq k, (q_h, \alpha_l, q_j) \in \delta} P(\mathbf{e}_l, \Gamma_j) \right).$$

This recurrence relation can be derived by observing that every path $\pi$ of length $i$ from $q_0$ to $q_j$ can be decomposed into the path $\pi[0, i-1]$ ending at some state $q_h$ and the path $\pi[i-1, i] = q_h \alpha_l q_j$. The cycle types $\Gamma_j$ can be "used" since $q_j$ is visited. The proof is in the appendix.

To finish the proof of Theorem 4.1, it suffices to give an algorithm with running time $2^{O(k^2 \log(kn))}$ for computing a desired semilinear basis $P_{i,j}$ for each set $T_{i,j}$. The algorithm runs in $(n-1)^2 + 1$ stages, where at stage $j = 0, \ldots, (n-1)^2$ the set $P_{i,j}$ is computed. Obviously, we first set $P_{0,0} = \{\mathbf{0}\}$ and $P_{0,j} = \emptyset$ for each $j \in \{1, \ldots, n-1\}$. Inductively, suppose that $P_{i,h} = \{\langle \mathbf{v}_s^h; S_s^h \rangle\}_{s=1}^{m_h}$ has been computed for each $h \in \{0, \ldots, n-1\}$. We will show how to compute $P_{i+1,j}$ for any given $j \in \{0, \ldots, n-1\}$. For each $h \in \{0, \ldots, n-1\}$, let $J_h$ denote the set of numbers $l \in \{1, \ldots, k\}$ such that $(q_h, \alpha_l, q_j) \in \delta$. Therefore, we have $P_{i,h} + \bigcup_{l \in J_h} P(\mathbf{e}_l, \Gamma_j) = \bigcup_{s=1}^{m_h} \bigcup_{l \in J_h} P(\mathbf{v}_s^h + \mathbf{e}_l, S_s^h \cup \Gamma_j)$. We use the algorithm from Theorem 3.1 to compute another semilinear basis for each $P(\mathbf{v}_s^h + \mathbf{e}_l, S_s^h \cup \Gamma_j)$ and then compute unions in the obvious way to obtain $P_{i+1,j}$ (note: duplicate is removed). The output of this algorithm is $P = \bigcup_{i=1}^{(n-1)^2} P_{i,n-1}$. The correctness of the algorithm is immediate from Lemma 4.6.

We now analyze the time complexity of this algorithm. By induction, it is easy to see that at every stage of the algorithm $S_s^h \cup \Gamma_j \subseteq \{0, \ldots, n\}^k$ holds for each $h \in \{0, \ldots, n-1\}$ and $s \in \{1, \ldots, m_h\}$. Therefore, the maximum component over all offsets in the semilinear basis $P_{i+1,j}$ is at most $a + O(n^{3(k+1)} k^{4k+6})$, where $a$ is the maximum entry in each $\mathbf{v}_s^h + \mathbf{e}_l$ over all $h \in \{0, \ldots, n-1\}$, $l \in J_h$, and $s \in \{1, \ldots, m_h\}$. Note that the summand $O(n^{3(k+1)} k^{4k+6})$ is due to an application of Theorem 3.1. By induction, at stage $i$ the maximum component over all offsets in $\{P_{i,j}\}_{j=0}^{n-1}$ is $i \times O(n^{3(k+1)} k^{4k+6})$. This means that the maximum entry of each offset in $P$ is $O(n^{3k+5} k^{4k+6})$, and the number of linear bases in $P$ is $O(n^{k^2+3k+5} k^{4k+6})$ (since duplicates are always removed). It is also easy to see that at each stage $i$, the algorithm runs in time $2^{O(k^2 \log(nk))}$, primarily spent in the algorithm from Theorem 3.1. All in all, our algorithm runs in time $2^{O(k^2 \log(nk))}$, which is also the complexity of the entire procedure.

## 5 Lower bounds

In this section, we shall prove three lower bounds to complement the results in the previous section. We start by proving that *every* semilinear basis for the Parikh image of a deterministic finite automaton (DFA) can be large in the size of the alphabet.

**Proposition 5.1** *For each* $k \in \mathbb{Z}_{>0}$ *and each integer* $n > 1$*, there exists a DFA* $A_{n,k}$ *over the alphabet* $\Sigma_k :=$ $\{a_1, \ldots, a_k\}$ with $n+1$ states whose Parikh image contains at least $n^{k-1}/(k-1)!$ linear sets.

**Proof.** Let $A_{n,k} = (Q = \{q_0, \ldots, q_n\}, \delta, q_0, q_n)$ with $\delta(q_i, a) = q_{i+1}$ for each $a \in \Sigma_k$ and $0 \leq i < n$. This automaton has a finite language $L(A_{n,k})$ with Parikh image $\mathcal{P}(L(A_{n,k}))$ containing precisely all *ordered* integer partitions of $n$ into $k$ parts. Since the set $\mathcal{P}(L(A_{n,k}))$ is finite, each ordered integer partition $(n_1, \ldots, n_k)$ of $n$ must appear in precisely one linear set. Finally, it is easy to check (e.g. see [31, Chapter 13]) that the number of ordered partitions of $n$ into $k$ parts equals $\binom{n+k-1}{k-1} \geq n^{k-1}/(k-1)!$. □

This proposition implies that, for every fixed $k \geq 1$, there exists infinitely many DFAs $\{A_n\}$ over an alphabet of size $k$ where $A_n$ has size $O(n)$ but $\mathcal{P}(A_n)$ must contain $\Omega(n^{k-1})$ linear bases. Therefore, this shows that $k$ *cannot* be removed from the exponent in Theorem 4.1. In addition, observing that the DFAs that we constructed have equivalent regular expressions of size $O(n)$, Proposition 5.1 also gives lower bounds for Parikh images of regular expressions.

Next, we show that Theorem 4.1 *cannot* be extended to CFGs (equivalently, pushdown automata). More precisely, we show that the number of linear sets for Parikh images of CFGs could be exponential in the size of the CFGs.

**Proposition 5.2** *There exists a small constant $c \in \mathbb{Z}_{>0}$ such that, for each integer $n > 1$, there exists a CFG $G_n$ of size at most $cn$ over the alphabet $\Sigma := \{a\}$ whose Parikh image contains precisely $2^n$ linear sets.*

**Proof.** We will construct a CFG $G_n$ such that $\mathcal{P}(L(G_n)) = \{0, 1, \ldots, 2^n - 1\}$, each of whose elements will appear in precisely one linear set. Our construction uses the lower bound technique in [24].

Our CFG $G_n$ contains nonterminals $S, \{A_i\}_{i=0}^{n-1}$, and $\{B_i\}_{i=0}^{n-1}$, and consists precisely of the following rules:

$$\begin{aligned} S &\to A_0 \ldots A_{n-1} \\ A_i &\to \varepsilon \quad \text{for each } 0 \leq i < n \\ A_i &\to B_i \quad \text{for each } 0 \leq i < n \\ B_i &\to B_{i-1} B_{i-1} \quad \text{for each } 0 < i < n \\ B_0 &\to a \end{aligned}$$

The initial nonterminal is declared to be $S$. It is easy to prove by induction that, for each word $w \in \Sigma^*$, $B_i \Rightarrow^* w$ iff $w = a^{2^i}$. This implies that $A_i$ generates either $\varepsilon$ or $a^{2^i}$. Thus, we see that $L(G_n) = \{a^i : 0 \leq i < 2^n\}$, which easily yields the desired result. □

Finally, we give a lower bound proving the tightness of quadratic upper bound in Proposition 4.3, even when restricted to DFAs.

**Proposition 5.3** *For each positive integer $n > 2$, there exists a DFA $A_n$ with $2n + 3$ states such that if $\mathcal{P}(L(A_n))$ can be represented as a union of the linear sets $P(\mathbf{v}_1; S_1), \ldots, P(\mathbf{v}_r; S_r)$, then one entry in some $\mathbf{v}_i$ is at least $n(n+1)/2$.*

This proof is given in the appendix. In fact, the constructed DFA $A_n$ has an equivalent regular expression of size $O(n)$ as well, therefore yielding the same quadratic lower bound for regular expressions.

## 6 Applications

**Integer Programming**

*Integer programming* (IP) is the problem of checking whether a given integer program $A\mathbf{x} = \mathbf{b}$ ($\mathbf{x} \succeq \mathbf{0}$), where $A$ is an $k$-by-$m$ integer matrix and $\mathbf{b} \in \mathbb{Z}^k$, has a solution. It is well-known that this problem is NP-complete [22]. On the other hand, for $k = 1$, there is a pseudopolynomial-time algorithm for this problem (a.k.a. knapsack problem), which remains NP-complete under binary representation of input numbers [22]. Furthermore, Papadimitriou [21] has established a pseudopolynomial-time algorithm for solving IP, for any fixed $k \geq 1$. The running time of his algorithm is $2^{O(k \log m + k^2 \log(ka))}$, where $a$ is the maximum absolute value of numbers appearing in the input. Using our results in the previous sections, we could show that the problem remains poly-time solvable (for any fixed $k \geq 1$) even if the numbers in $\mathbf{b}$ are given in *binary* (and $A$ in unary).

**Theorem 6.1** *Given a $k$-by-$m$ integer matrix $A$ and a vector $\mathbf{b} \in \mathbb{Z}^k$, let $a$ (resp. $b$) be the maximum absolute value of numbers appearing in $A$ (resp. $\mathbf{b}$). Then, checking whether the integer program $A\mathbf{x} = \mathbf{b}$ ($\mathbf{x} \succeq \mathbf{0}$) has a solution can be done in time $2^{O(k \log m + k^2 \log(ka) + \log \log b)}$.*

The proof of this theorem is in the appendix. It is essentially a simple application of Theorem 3.1 and Kannan's poly-time algorithm [17] for IP for a fixed number of variables, which improves the running time of Lenstra's original poly-time algorithm [19]. Its running time on an integer program with $d$ variables and input of length $L$ is $O(d^{9d} L \log L)$, which is polynomial even under binary encoding of numbers in input.

Theorem 6.1 is known to be true when $k = 1$, by solutions to the *Frobenius's problem*, and has such an application as the *coin problem* (e.g. see [25]). We give another not so obvious application of Theorem 6.1 in the appendix.

**Membership for Parikh images of NFAs**

The *membership problem for Parikh images of NFAs* is defined as follows: given an NFA $A$ over $\Sigma = \{\alpha_1, \ldots, \alpha_k\}$

and a tuple $\mathbf{b} \in \mathbb{N}^k$ given in binary, decide whether $\mathbf{b} \in \mathcal{P}(L(A))$. A similar membership problem could be defined for CFGs (equivalently, pushdown automata). This problem is known to be NP-complete (e.g. see [10]). Notice, however, that the hardness proof in [10] requires $k$ to be unbounded. Furthermore, since it is well-known that NFAs can be efficiently converted to equivalent pushdown autommata (equivalently, CFGs), it follows that the membership problem for Parikh images of NFAs is in NP. [See [26, 32] for an alternative proof via existential Presburger arithmetic.] We first give two complementary lower bounds for these results.

**Proposition 6.2** *Membership for Parikh image of NFAs (resp. CFGs) is NP-hard when $k$ is not fixed (resp. even when $k = 1$).*

NP-hardness already holds for DFAs (reduction from the hamiltonian path problem) and for regular expressions (reduction from one-in-three 3SAT). For CFGs over unary alphabet, we use the encoding of large numbers given in Proposition 5.2 to encode the 0-1 knapsack problem.

This leaves the complexity of the membership problem for Parikh images of NFAs for any fixed $k \geq 1$. Combining Kannan's poly-time algorithm for IP over a fixed number of variables (as in the previous application) and Theorem 4.1, it is easy to show that this problem is solvable in time polynomial in $|A|$ and $\|\mathbf{b}\|$ but exponential in $k$, i.e., polynomial when $k$ is fixed.

**Theorem 6.3** *Given an NFA $A$ with $n$ states over the alphabet $\Sigma = \{\alpha_1, \ldots, \alpha_k\}$ and a tuple $\mathbf{b} = (b_1, \ldots, b_k) \in \mathbb{N}^k$ written in binary with $b := \max_{1 \leq i \leq k}\{b_i\}$, checking whether $\mathbf{b} \in \mathcal{P}(L(A))$ can be done in time $2^{O(k^2 \log(kn) + \log \log b)}$.*

**Polynomial PAC-learnability of semilinear sets**

Valiant's notion of PAC (*Probably Approximately Correct*) learning is a standard model in computational learning theory [3, 4]. In this framework, a learning algorithm is required to run in time polynomial in the size of the training sample, and output a hypothesis for an (unknown) target concept that is as *precise* as the "user" desires, given any sufficiently large training sample (but still polynomial in the reciprocals of approximation/confidence parameters).

The issues of PAC-learnability of semilinear sets have been addressed by Abe [1]. In particular, learning semilinear sets of dimension 1 under binary representation of numbers is shown to be as hard as learning boolean formulas in DNF, which is (still) a major open question in learning theory. On the positive side, he shows that semilinear sets of dimension 1 and 2 can be poly-time PAC-learned when the numbers are represented in unary. To this end, he established a *normal form lemma* for semilinear sets of dimension 2 (in unary), which is simply a special case of Theorem 3.1 for dimension 2. However, his proof makes use of geometric facts that are specific to $\mathbb{R}^2$. For this reason, he leaves open the learnability question of semilinear sets in unary over any fixed dimension $k > 2$ [1, Section 9]. Replacing Abe's normal form lemma with Theorem 3.1 and following the proof of [1, Theorem 6.1], we can easily deduce the more general theorem.

**Theorem 6.4** *Semilinear sets in unary representation over any fixed dimension $k \geq 1$ can be polynomially PAC-learned with respect to concept complexity.*

We shall not reproduce the proof for this theorem as it essentially requires only replacing Abe's normal form lemma with Theorem 3.1 in the proof of [1, Theorem 6.1]. A short sketch is provided in the appendix.

**Verifying counter systems**

Minsky's counter machines are well-known to be a Turing-powerful model of computation. In verification literature, many decidable subclasses of counter machines have been studied including *reversal-bounded* counter systems [15, 16, 30], and their extensions with pushdown stacks and discrete clocks [7, 30]. Intuitively, $r$-reversal $k$-counter systems are simply Minsky's counter machines with $k$ counters, each of which can change from an incrementing mode to a decrementing mode (or vice versa) only for a fixed $r$ number of times. Their connection to our result is due to the use of Parikh's Theorem for obtaining decidability (initially used in [16]). Due to space restriction, we shall briefly mention only one corollary of our result.

In [30], it was shown that model checking Linear Temporal Logic (LTL) over reversal-bounded counter systems with discrete clocks is solvable in double exponential time, even when one of the counters is *free* (not reversal-bounded). More precisely, if $t$ is the number of clocks, $n$ the number of states in the finite control, $c$ the size (in binary) of the maximum number appearing in clock constraints, and $\|\varphi\|$ the size of the input LTL formula $\varphi$, then model checking LTL on such systems is decidable in time exponential in $n$ but double exponential in $c$, $k$, $r$, $\|\varphi\|$, and $t$. This result was derived by carefully analyzing the complexity of Ibarra's original result [16] of effective semilinearity of the Parikh images of languages recognized by reversal-bounded counter machines and replacing the application of Parikh's Theorem in [16] by the linear translation of [32] from CFGs to existential Presburger formulas expressing their Parikh images. By replacing the application of the translation of [32] by Theorem 4.1 and not allowing a single free counter, we can easily obtain the following upper bound.

**Theorem 6.5** *Model checking LTL over reversal-bounded counter systems with discrete clocks is solvable in time polynomial in $n$ and exponential in $c$, $k$, $r$, $\|\varphi\|$, and $t$.*

We shall not reproduce the proof (rather see the full version of the paper [30], which can be requested from the author). The time complexity given by this theorem is tight for the following reason: 1) LTL model checking over finite systems is already PSPACE-complete [28], 2) emptiness for $r$-reversal $k$-counter automata is PSPACE-hard [15], 3) emptiness for discrete timed automata is PSPACE-complete even when restricted to three clocks *or* when $c$ is "small" (see [2] and references therein).

# References

[1] N. Abe. Characterizing PAC-Learnability of Semilinear Sets. *Inf. Comput.* 116(1): 81-102 (1995)

[2] R. Alur and P. Madhusudan. Decision Problems for Timed Automata: A Survey. In *SFM'04*, pages 1–24.

[3] D. Angluin. Computational learning theory: survey and selected bibliography. In *STOC'92*, pages 351–369.

[4] M. Anthony and N. Biggs. *Computational learning theory: an introduction.* Cambridge University Press, 1992.

[5] M. Bojanczyk. A new algorithm for testing if a regular language is locally threshold testable. *IPL* 104(3):91–94 (2007)

[6] M. Chrobak. Finite automata and unary languages. In *TCS* 302 (2003) 497–498.

[7] Z. Dang, O. Ibarra, T. Bultan, R. Kemmerer, J. Su. Binary reachability analysis of discrete pushdown timed automata. In *CAV'00*, pages 69–84.

[8] L. E. Dickson. Finiteness of the odd perfect and primitive abundant numbers with $r$ distinct prime factors. *Amer. Journal Math.*, 35:413–422, 1913.

[9] J. Edmonds. Systems of distinct representatives and linear algebra. *J. Res. Nat. Bur. Standards* 71B:241–245 (1967)

[10] J. Esparza. Petri nets, Commutative Context-Free Grammars, and Basic Parallel Processes. *Fundam. Inform.* 31(1):13–25 (1997)

[11] W. Gelade and F. Neven. Succinctness of the Complement and Intersection of Regular Expressions. In *STACS'08*, pages 325–336.

[12] S. Ginsburg and E. H. Spanier. Semigroups, Presburger Formulas, and Languages. *Pacific Journal of Mathematics* 16(2):285–296 (1966)

[13] S. Göller, R. Mayr and A. W. To. On the Computational Complexity of Verifying One-Counter Processes. In *LICS'09*, pages 235-244.

[14] H. Gruber and M. Holzer. Computational complexity of NFA minimization for finite and unary languages. In LATA'08, pages 261–272.

[15] E. M. Gurari and O. H. Ibarra. The Complexity of Decision Problems for Finite-Turn Multicounter Machines. *JCSS* 22:220–229 (1981)

[16] O. Ibarra. Reversal-bounded multicounter machines and their decision problems. *J. ACM* 25 (1978), 116–133.

[17] R. Kannan. Improved algorithms for integer programming and related lattice problems. In *STOC'83*, pages 193–206.

[18] D. Kozen. *Automata and Computability*. Springer, 1997.

[19] H. W. Lenstra. Integer programming with a fixed number of variables. *Math. Oper. Res.*, 8(4):538–548 (1983)

[20] A. Martinez. Efficient computation of regular expressions from unary NFAs. In *DFCS'02*, pages 174–187.

[21] C. H. Papadimitriou. On the complexity of integer programming. *J. ACM* 28(4): 765–768 (1981)

[22] C. H. Papadimitriou and K. Steiglitz. *Combinatorial Optimization: Algorithms and Complexity* Dover Publications, 1998.

[23] R. Parikh. On Context-Free Languages. *J. ACM* 13(4):570–581 (1966)

[24] G. Pighizzini, J. Shallit, and M. Wang. Unary Context-Free Grammars and Pushdown Automata, Descriptional Complexity and Auxiliary Space Lower Bounds. *JCSS* 65(2): 393–414 (2002)

[25] J. L. Ramírez Alfonsín. *The Diophantine Frobenius Problem*. Oxford University Press, 2005.

[26] H. Seidl, Th. Schwentick, and A. Muscholl. Counting in Trees. *Texts in Logic and Games* 2:575–612, 2007.

[27] M. Sipser. *Introduction to the Theory of Computation*. PWS Publishing Company, 1997.

[28] A. P. Sistla and E. M. Clarke. The Complexity of Propositional Linear Temporal Logics. *JACM* 32(3):733–749 (1985)

[29] A. W. To. Unary finite automata vs. arithmetic progressions. *IPL* 109(17): 1010–1014 (2009)

[30] A. W. To and L. Libkin. Algorithmic metatheorems for decidable LTL model checking over infinite systems. To appear in *FoSSaCS '10*.

[31] J. H. van Lint and R. M. Wilson. *A Course in Combinatorics*. (second ed.) Cambridge University Press, 2001.

[32] K. N. Verma, H. Seidl, T. Schwentick. On the complexity of equational Horn clauses. In *CADE'05*, pages 337–352.

[33] G. M. Ziegler. *Lectures on Polytopes*. Springer, 2007.

# APPENDIX

## A Proof of Lemma 3.2

The conical version of Caratheodory's theorem [33, Proposition 1.15] states that if we take $V_1, \ldots, V_s$ to be the sequence of all subsets of $V$ with $|V_i| \leq \textbf{rank}(V) \leq k$, then we have $\textbf{cone}(V) = \bigcup_{i=1}^{s} \textbf{cone}(V_i)$. Let $d := \textbf{rank}(V)$ and $S_1, \ldots, S_r$ be the set of all distinct linearly independent $d$-subsets of $V$. It is not difficult to see that $\bigcup_{i=1}^{s} \textbf{cone}(V_i) = \bigcup_{i=1}^{r} \textbf{cone}(S_i)$. In fact, we obviously have $\bigcup_{i=1}^{s} \textbf{cone}(V_i) \supseteq \bigcup_{i=1}^{r} \textbf{cone}(S_i)$. Conversely, observe that for each linearly independent subset $V_i$ of rank $d' < d$, one could add a vector $\mathbf{v} \in V \setminus V_i$ to $V_i$ yielding a linearly independent $(d'+1)$-subset $V_i \cup \{\mathbf{v}\}$ of $V$; otherwise, each vector $\mathbf{v} \in V \setminus V_i$ is a linear combination of vectors in $V_i$ contradicting that $d' < d$. Repeating this process several times, we obtain a linearly independent $d$-subset of $V$ containing $V_i$. This easily implies that $\bigcup_{i=1}^{s} \textbf{cone}(V_i) \subseteq \bigcup_{i=1}^{r} \textbf{cone}(S_i)$.

Observe now that $r \leq \binom{m}{d} \leq m^d \leq m^k$. So, we first compute the rank $d$ of $V$ by Gaussian-elimination in the standard way. It is known [9] that Gaussian-elimination over rational numbers can be implemented to run in time polynomial in the total number of bits in the input matrix. We then obtain $S_1, \ldots, S_r$ by going through each $d$-subset of $V$ and using Gaussian-elimination to check whether it has rank $d$. In total, the running time is $2^{O(d \log m + \log(kma))} = 2^{O(k \log m + \log a)}$.

## B Proof of Lemma 3.3

That $\trianglelefteq_j$ is a partial order is due to:

- Observations **(O1)** and **(O2)**: the uniqueness of choice of canonical vector $\mathbf{v}_0$ and coefficients $a_1, \ldots, a_d \in \mathbb{N}$ for each vector $\mathbf{v}$ satisfying $\mathbf{v} = \mathbf{v}_0 + \Sigma_{i=1}^{d} a_i \mathbf{u}_i$.

- That $\preceq$ is a partial order on $\mathbb{N}^k$.

To see that $\trianglelefteq_j$ is well-founded, assume that there exists a strictly decreasing sequence $\mathbf{v}_1 \triangleright_j \mathbf{v}_2 \triangleright_j \ldots$. Let $\mathbf{v}_i = \mathbf{v}_0 + \Sigma_{j=1}^{d} a_j^i \mathbf{u}_i$ for some unique canonical vector $\mathbf{v}_0 = [\mathbf{v}_i]$ and unique coefficients $\mathbf{a}^i = (a_1^i, \ldots, a_d^j) \in \mathbb{N}^d$. In this way, we generate a strictly decreasing sequence $\mathbf{a}^1 \succ \mathbf{a}^2 \succ \ldots$ for the well-founded partial order $\succ$ on $\mathbb{N}^k$, and therefore a contradiction. Thus, $\trianglelefteq_j$ is a well-founded partial order on $\textbf{cone}_\mathbb{N}(V) \cap \textbf{cone}(S_j)$.

Given a $\trianglelefteq_j$-minimal vector $\mathbf{v} \in \textbf{cone}_\mathbb{N}(V) \cap \textbf{cone}(S_j)$, it is obvious that none of the vectors $(\mathbf{v} - \mathbf{u}_1), \ldots, (\mathbf{v} - \mathbf{u}_d)$ cannot be in $\textbf{cone}_\mathbb{N}(V) \cap \textbf{cone}(S_j)$. Conversely, given a vector $\mathbf{v} \in \textbf{cone}_\mathbb{N}(V) \cap \textbf{cone}(S_j)$ which is not $\trianglelefteq_j$-minimal, we could find another vector $\mathbf{v}' \in \textbf{cone}_\mathbb{N}(V) \cap \textbf{cone}(S_j)$ such that $\mathbf{v}' \triangleleft_j \mathbf{v}$. Using Observation **(O3)**, it is easy to show that at least one of the vectors $(\mathbf{v} - \mathbf{u}_1), \ldots, (\mathbf{v} - \mathbf{u}_d)$ is in $\textbf{cone}_\mathbb{N}(V) \cap \textbf{cone}(S_j)$.

## C Proof of Lemma 3.5

That $\mathbf{w}$ is a solution is immediate. To show $\preceq$-minimality, consider a vector $\mathbf{u} = (c_1', \ldots, c_m', b_1', \ldots, b_d') \in \mathbb{N}^{m+d}$ such that $\mathbf{u} \preceq \mathbf{w}$ and $A\mathbf{u}^\mathsf{T} = \mathbf{v}_0$. Define $\mathbf{v}' := \mathbf{v}_0 + \Sigma_{i=1}^{d} b_i' \mathbf{u}_i$ and thus $\mathbf{v}' = \Sigma_{i=1}^{m} c_i' \mathbf{v}_i$. This means that $\mathbf{v}' \in \textbf{cone}_\mathbb{N}(V) \cap \textbf{cone}(S_j)$ and, by $\trianglelefteq_j$-minimality of $\mathbf{v}$, it follows that $\mathbf{v}' = \mathbf{v}$ and thus $b_i' = b_i$ for every $1 \leq i \leq d$. That $\mathbf{c}$ is a $\preceq$-minimal solution to the integer program $\Sigma_{i=1}^{m} x_i \mathbf{v}_i = \mathbf{v}$ ($\mathbf{v} \succeq \mathbf{0}$) implies that $c_i' = c_i$ for every $1 \leq i \leq m$ and, thus, $\mathbf{u} = \mathbf{w}$.

## D Proof of Lemma 4.4

We only show that if the $M_\mathbf{v}[j, h] = 1$, then so is the $(j, h)$-component of the matrix in the r.h.s. The converse can be proved by observing that all the steps below can be easily reversed.

Let $s = 1 + \Sigma_{i=1}^{k} r_i$. Suppose that $\pi = q_{l_0} \beta_1 q_{l_1} \ldots \beta_s q_{l_s}$ is a path from $q_j$ to $q_h$ with $\mathcal{P}(\pi) = \mathbf{v}$. Thus, we have $l_0 = j$ and $l_s = h$. We now decompose $\pi$ as follows. Let $t$ be the first position where the letter $\alpha_i$ occurs in $\pi$, i.e., $\beta_t = \alpha_i$ and $\beta_{t'} \neq \alpha_i$ for all $t' < t$. Let $\pi_1 := q_{l_0} \beta_1 \ldots q_{l_{t-1}}$, $\pi_2 := q_{l_{t-1}} \beta_t q_{l_t}$, and $\pi_3 := q_{l_t} \beta_{t+1} \ldots q_{l_s}$. Let $\mathbf{u} := \mathcal{P}(\pi_1)$ and $\mathbf{w} := \mathcal{P}(\pi_3)$. Notice that the $i$th entry of $\mathbf{u}$ is 0 and $\mathbf{w} = \mathbf{v} - \mathbf{e}_i - \mathbf{u}$. Furthermore, we have $M_\mathbf{u}[j, l_{t-1}] = M_{\mathbf{e}_i}[l_{t-1}, l_t] = M_\mathbf{w}[l_t, h] = 1$. It follows that the $(j, h)$-component of the matrix $M_\mathbf{u} \bullet M_{\mathbf{e}_i} \bullet M_\mathbf{w}$ is 1.

## E Proof of Lemma 4.5

This can be done using Lemma 4.4 and dynamic programming. The algorithm has $n + 1$ stages. At stage $j = 0, \ldots, n$, we compute all $M_\mathbf{v}$ where the components in $\mathbf{v}$ sum up to $j$. These will be saved in the memory for subsequent stages of the iteration. As base cases, we would obtain $M_\mathbf{0}$ and $M_{\mathbf{e}_i}$, for each $1 \leq i \leq k$, directly from the input. Notice that boolean matrix multiplication can be done in $O(n^3)$ and so each stage of the computation can be performed in $O(n^{2k+4})$. Thus, the entire computation runs in time $2^{O(k \log n)}$.

## F Proof of Lemma 4.6

We first define the following notation: for every path $\pi$ in $A$, let $\Gamma_\pi$ denote the union of all $\Gamma_h$ over all $h \in \{0, \ldots, n-1\}$ such that $q_h$ occurs in $\pi$.

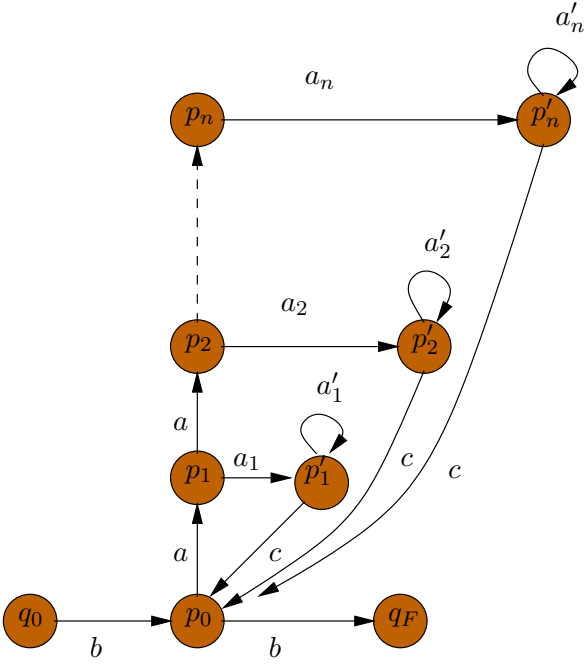

**Figure 1. A depiction of the DFA $A_n$**

($\subseteq$) Let $\mathbf{v} \in T_{i,j}$. Then, there exists a path $\pi = q_{l_0}\beta_1 \ldots \beta_i q_{l_i}$ such that $\mathbf{v} \in T_\pi$, where $l_0 = 0$ and $l_i = j$. Let $\pi'$ be the path segment $\pi[0, i-1]$ of $\pi$. Thus, we have $\Gamma_{\pi'} \subseteq \Gamma_\pi$. Suppose that $\Gamma_\pi = \{\mathbf{u}_1, \ldots, \mathbf{u}_m\}$ and $\Gamma_{\pi'} = \{\mathbf{u}_1, \ldots, \mathbf{u}_{m'}\}$ for some $m' \leq m$. It follows that $\Gamma_{\pi_w} - \Gamma_{\pi_{w'}} \subseteq \Gamma_j$. In other words, new cycle type could only be introduced by visiting $q_j$. Since $\mathbf{v} \in T_\pi$, we have $\mathbf{v} = \mathcal{P}(\pi) + \Sigma_{s=1}^m a_s \mathbf{u}_s$ for some $a_1, \ldots, a_m \in \mathbb{N}$. Thus, if $\alpha_l = \beta_i$, it follows that $\mathbf{v} = (\mathcal{P}(\pi') + \Sigma_{s=1}^{m'} a_s \mathbf{u}_s) + (\mathbf{e}_l + \Sigma_{s=m'+1}^m a_s \mathbf{u}_s)$. This establishes that $\mathbf{v} \in T_{i,h} + P(\mathbf{e}_l, \Gamma_j)$ as desired.

($\supseteq$) This inclusion could be proved by observing that all the steps in the converse case could be easily reversed.

## G  Proof of Proposition 5.3

The automaton $A_n = (\Sigma_n, Q_n, \delta_n, q_0, q_F)$, where $Q_n = \{q_0, q_F\} \cup \{p_0, \ldots, p_n\} \cup \{p'_1, \ldots, p'_n\}$ and $\Sigma_n = \{a, b, c\} \cup \{a_i, a'_i : 1 \leq i \leq n\}$. We specify the transition function $\delta_n$ in Figure 1. Notice that $A_n$ has an equivalent regular expression $R_n$ of size $O(n)$. For example, when $n = 2$, we can define $R_n$ to be

$$b(a(a_1(a'_1)^*c|aa_2(a'_2)^*c))^*b.$$

We now argue that the $a$-component of some $\mathbf{v}_i$ must be at least $n(n+1)/2$. Let $m$ be the maximum entry over all vectors in $\bigcup_{i=1}^r S_i$. Define $N :=$ $\left(\max\{|S_i| : 1 \leq i \leq r\}\frac{n(n+1)}{2}m\right) + 1$. For each $1 \leq i \leq n$, let $C_i$ be the cycle $p_0 a p_1 a \ldots p_i a_i p'_i (a'_i p'_i)^N c p_0$. Consider the accepting path $\pi = (q_0 b p_0) \odot C_1 \odot C_2 \odot \cdots \odot C_n \odot (p_0 b q_F)$. We have $\mathcal{P}(\pi) \in P(\mathbf{v}_h; S_h)$ for some $1 \leq h \leq r$. Observe also that $a$ occurs precisely $\Sigma_{i=1}^n i = n(n+1)/2$ times in $\pi$.

**Claim.** Each $a'_i$-component of $\mathbf{v}_h$ ($1 \leq i \leq n$) is nonzero.

We now prove this claim. Let $S_h = \{\mathbf{u}_1, \ldots, \mathbf{u}_s\}$ and $\mathcal{P}(\pi) = \mathbf{v}_h + \Sigma_{i=1}^s t_i \mathbf{u}_i$. For each $i \in \{1, \ldots, n\}$, there exists a vector $\mathbf{u}_{j_i}$ with a positive $a'_i$-component and $t_{j_i} > n(n+1)/2$; for, otherwise, the $a'_j$-component of $\mathcal{P}(\pi)$ is at most $|S_h|\frac{n(n+1)}{2}m < N$, a contradiction. In particular, this implies that all $\alpha$-component of $\mathbf{u}_{j_i}$, where $\alpha \neq a'_i$ for all $i \in \{1, \ldots, n\}$, is 0 as each such letter $\alpha$ occurs at most $n(n+1)/2$ times in $\pi$. But this means that each $a_i$-component of $\mathbf{v}_h$ is nonzero; for, otherwise, we could consider the vector $\mathbf{v}_h + \mathbf{u}_{j_i}$ which would not correspond to any accepting path in $A_n$ since at least one letter $a_i$ needs to read by $A_n$ if $a'_i$ is to occur in the path. This proves our claim.

Now consider each word $w = w_0 \ldots w_l \in L(A_n)$ such that $\mathcal{P}(w) = \mathbf{v}_h$. It is easy to see that the number of occurences of $a$ in $w$ must be at least $n(n+1)/2$. In fact, for every $1 \leq i \leq n$, define $j_i = \min\{j : w_j = a_i\}$. For each $i$, the number of occurences of $a$ in $w_t \ldots w_{j_i}$, where $t = \max\{j_{i'} : j_{i'} < j_i, 1 \leq i' \leq n\}$, is at least $i$. The lower bound of $n(n+1)/2$ on the number of occurences of $a$ in $w$ immediately follows.

## H  Proof of Theorem 6.1

To prove this theorem, let $\mathbf{v}_i$ ($1 \leq i \leq m$) be the $i$-th column vector of $A$ and write $\mathbf{x}^T = (x_1, \ldots, x_m)$. Then, we have $A\mathbf{x} = \Sigma_{i=1}^m x_i \mathbf{v}_i$ and thus checking for solvability of $A\mathbf{x} = \mathbf{b}$ is equivalent to checking whether $\mathbf{b} \in \mathbf{cone}_\mathbb{N}(V)$, where $V = \{\mathbf{v}_1, \ldots, \mathbf{v}_m\}$. By Theorem 3.1, we obtain in time $2^{O(k \log m + k^2 \log(ka))}$ some semi-linear sets $P(\mathbf{w}_1; S_1), \ldots, P(\mathbf{w}_r; S_r)$ with $\mathbf{cone}_\mathbb{N}(V) = \bigcup_{i=1}^s P(\mathbf{w}_i; S_i)$, where the maximum absolute value of entries of each $w_i$ is $O(m(k^2a)^{2k+3})$, each $S_i$ is a subset of $V$ with $|S_i| \leq k$, and $s = O(m^{k+1}(k^2a)^{2k+3})$. We would then simply check whether $\mathbf{b} \in P(\mathbf{w}_i; S_i)$ for some $i = 1, \ldots, s$. To this end, we employ Kannan's polynomial-time algorithm [17] for IP for fixed number of variables, which improves the running time of Lenstra's original polynomial-time algorithm [19]. Its running time on an integer program with $d$ variables and input of length $L$ is $O(d^{9d}L \log L)$. Observe now that each number in $S_i$ and $\mathbf{w}_i$ could be represented using $O(k \log(mak))$ bits. Therefore, if $S_i =$ $\{\mathbf{u}_1, \ldots, \mathbf{u}_d\}$, the integer program $\Sigma_{i=1}^d x_i \mathbf{u}_i = \mathbf{b} - \mathbf{w}_i$ has length $L = O(k \log b + k^3 \log(mak))$ and uses $d \leq k$ variables. On this input, Kannan's algorithm will terminate in time $2^{O(k \log k + \log \log b + \log \log(ma))}$. In the worst case, we will need to run Kannan's algorithm for each $i = 1, \ldots, s$ if the answer to the original IP question is negative. In total, the algorithm runs in time $2^{O(k \log m + k^2 \log(ka) + \log \log b)}$.

## I Another application of Theorem 6.1

Another not immediately obvious application is the following *navigation problem*. Suppose we have a robot in $\mathbb{Z}^k$ that can make finitely many different *types* of local moves $\mathbf{v}_1, \ldots, \mathbf{v}_m \in \mathbb{Z}^k$ at any given point in $\mathbb{Z}^k$. It is reasonable to assume that local moves are small (i.e. given in unary). We are interested in the following global behavior: when initially placed at $\mathbf{0}$, can it eventually arrive at a given vector $\mathbf{b}$? The vector $\mathbf{b}$ can be assumed to be large (i.e. given in binary) as it pertains to the result of a global behavior. Theorem 6.1 implies that this navigation problem is poly-time solvable when the dimension $k$ is fixed.

## J Proof of Proposition 6.2

**DFA** We now give a poly-time reduction from the hamiltonian path problem to membership problem for Parikh images of DFAs. The hamiltonian path problem asks whether a given graph $G = (V = \{v_1, \ldots, v_n\}, E)$ with a source $s := v_1 \in V$ and a target $t := v_n \in V$ has a hamiltonian path from $s$ to $t$, i.e., a path from $s$ to $t$ in $G$ that visits each vertices in $V$ *exactly* once. Given $G, s, t$, we define the DFA $A_{G,s,t} = (\Sigma, Q, \delta, q_0, q_F)$ where $Q := V$, $\Sigma := \{\alpha_1, \ldots, \alpha_n\}$, $q_0 := s$, $q_F := t$, and $\delta := \{(v_i, \alpha_j, v_j) : (v_i, v_j) \in E\}$. Then, it is easy to see that $G$ has a hamiltonian path from $s$ to $t$ iff the Parikh image $\mathcal{P}(\alpha_1 \ldots \alpha_n)$ of the word $\alpha_1 \ldots \alpha_n$ is in $\mathcal{P}(A_{G,s,t})$ iff $(0, 1, 1, \ldots, 1) \in \mathcal{P}(A_{G,s,t})$. This completes the proof of NP-hardness of the membership problem for Parikh images of DFAs with unbounded alphabet size.

**Regular expressions** One-in-three 3SAT is the following problem: given a boolean formula $\varphi$ in 3-CNF, does there exist a satisfying assignment for $\varphi$ that additionally makes no more than one literal true for each clause. We shall call such an satisfying assignment *1-in-3*. This problem is NP-complete (see "M. R. Garey and D. S. Johnson. *Computers and Intractability: A Guide to the Theory of NP-completeness*, W. H. Freeman, 1979."). We shall reduce this problem to the membership problem for Parikh images of regular expressions. Given $\varphi = C_1 \wedge \ldots \wedge C_k$, where $C_i$ is a multiset over $L := \{x_1, \bar{x}_1 \ldots, x_n, \bar{x}_n\}$ with $|C_i| = 3$, we define a function $f : L \to \{1, \ldots, k\}^*$ as follows:

- $f(x_i) := a_1 \ldots a_k$ where $a_i := i$ if $x_i \in C_j$, and $a_i := \varepsilon$ otherwise.

- $f(\bar{x}_i) := a_1 \ldots a_k$ where $a_i := i$ if $\bar{x}_i \in C_j$, and $a_i := \varepsilon$ otherwise.

That is, the function $f$ associates a literal with the indices of clauses that are satisfied when the value 1 is assigned to the literal. The corresponding regular expression over $\Sigma = \{1, \ldots, k\}$ is

$$R_\varphi = (f(x_1)|f(\bar{x}_1)) \ldots (f(x_n)|f(\bar{x}_n)).$$

Let $\mathbf{1} \in \{1\}^k$. We claim that $\varphi$ is a positive instance of one-in-three 3SAT iff $\mathbf{1} \in \mathcal{P}(L(R_\varphi))$. To prove this, suppose that $\varphi$ is a positive instance with a 1-in-3 satisfying assignment $\sigma : L \to \{0, 1\}$ (i.e. $\sigma(x_i) = 1$ iff $\sigma(\bar{x}_i) = 0$). Consider the word $w := X_1 \ldots X_n \in \Sigma^*$, where

$$X_i := \begin{cases} f(x_i) & \text{if } \sigma(x_i) = 1, \\ f(\bar{x}_i) & \text{if } \sigma(x_i) = 0. \end{cases}$$

Observe that $w \in L(R_\varphi)$. Since $\sigma$ is a 1-in-3 satisfying assignment, it follows that $\mathcal{P}(w) = \mathbf{1}$ and, therefore, we have $\mathbf{1} \in \mathcal{P}(L(R_\varphi))$. The converse direction can be proved by reversing the above construction of the word $w$. Finally, observe that the construction of $R_\varphi$ and $\mathbf{1}$ can be done in time polynomial in the size of $\varphi$.

**CFG** For CFGs over unary alphabet, we could easily use the encoding of large numbers given in Proposition 5.2 to succinctly encode the NP-complete 0-1 knapsack problem, i.e., given $a_1, \ldots, a_m \in \mathbb{N}$ and $b \in \mathbb{N}$ all represented in binary, decide whether there exists a subset $S \subseteq \{1, \ldots, m\}$ such that $\Sigma_{i \in S} a_i = b$.

## K Sketch for proving Theorem 6.4

The reader should first familiarize themselves with the notions of PAC-learnability. Abe [1, Section 2–4] gives a highly-readable exposition; but a more gentle introduction could be found in [3, 4]. In the following, a semilinear set is said to *have at most $k$ generators* if it is a union of linear sets with at most $k$ generators.

To prove Theorem 6.4, the reader should first recall the notion of *polynomially generable natural union class (PGNU)* in [1, Definition 6.1–2]. [Definition 6.2 has a typo in item 2(ii): $size(M'') < size(M)$ should be $size(M'') < size(M')$.] By [1, Lemma 6.1] and [1, Lemma 4.6], it suffices to show that for each fixed $k \geq 1$:

**(P1)** semilinear sets in unary representation with at most $k$ generators over any fixed dimension $k \geq 1$ is a PGNU (an adaptation of [1, Corollary 6.1]); and **(P2)** semilinear sets in unary with at most $k$ generators over $\mathbb{N}^k$ is a *Poly Blow-up Normal Form* (c.f. [1, Definition 4.9]) for semilinear sets in unary over $\mathbb{N}^k$ (an adaptation of [1, Corollary 6.2]).

The proof of **(P1)** is almost identical to the proof of [1, Corollary 6.1]:

1. for any given $b \in \mathbb{Z}_{\geq 1}$, there are at most $k(b+1)^{k(k+1)}$ semilinear sets over $\mathbb{N}^k$ with at most $k$ generators whose offsets and generators contain only numbers in $\{0, \ldots, b\}$; and

2. (**First condition of PGNU**) it is easy to see that checking whether a given tuple $(a_1, \ldots, a_k) \in \mathbb{N}^k$ is a member of a given semilinear set with at most $k$ generators over $\mathbb{N}^k$ can be poly-time (truth-table) reduced to integer programming with fixed number of variables, which is poly-time solvable by Kannan's algorithm [17] (in this case, numbers could even be represented in binary).

The next step is to show **(P2)**. However, this is simply a corollary of our Theorem 3.1. This completes the proof of Theorem 6.4.